# Human Extinction: A Demographic Perspective

David A. Swanson[1] and Jeff Tayman[2]

[1] Center for Studies in Demography and Ecology, University of Washington, [2] Tayman Demographics

**Note and acknowledgements**

1. David A. Swanson is Distinguished Professor Emeritus, University of California Riverside and is an affiliate of the Center for Studies in Demography and Ecology, University of Washington. Jeff Tayman is the retired Director for Research, San Diego Association of Governments and is based in San Diego, California.

2. Please direct all correspondence to Dr. David A. Swanson, Center for Studies in Demography and Ecology, 206 Raitt Hall, Box 353412, 1410 NE Campus Parkway, University of Washington, Seattle, WA 98195-3423 USA. Email: dswanson@ucr.edu

3. The authors received no financial support for this research.

4. Each author contributed approximately equally to each of the areas of work underlying this paper

**ORCID**

David A. Swanson, https://orcid.org/0000-0003-4284-9478

Jeff Tayman, https://orcid.org/0000-0003-3572-209X

## Abstract

Studies that predict species extinction have focused on a range of flora and fauna but in regard to Homo sapiens there are, with one notable exception, no predictive studies, only considerations of possible ways this may occur. The exception believes extinction of Homo sapiens will happen in 10,000 years. We agree that extinction will happen, but we disagree on the timing: The work we present here suggests that if the current decline in birth rates continues, humans could be extinct by 2394. If we consider the absence of working-age people and the accompanying collapse of services, the survivorship rates would most likely be lower. Given this, it is plausible that extinction could occur around 2359. We also examined a scenario in which births ended in 2024, which revealed that Homo sapiens would become extinct in 2134. Given societal collapse, extinction under the zero births scenario could occur around 2089.

**Introduction**

There are predictive studies on the extinction of a range of flora and fauna (e.g., Cooke, Eigenbrod, and Bates, 2019; Foster, et al., 2023; Jones, Zurrell, and Wiesner, 2023; Liu et al., 2024; Toussaint et al., 2021) but none on Homo sapiens. Those who do consider the extinction of our species, with one notable exception, look at it as a potential rather than a inevitable event (Bostrom, 2002, 2009, 2013; Bostrom and Ćirković, 2011; Ćirković, Sandberg, and Bostrom, 2010; Ehrlich, 1969; Lutz and Qiang, 2002; Meadows et al., 1972; Posner, 2004; Raup, 1986). The exception is Henry Gee (2025: 233), who not only believes human extinction is likely but that it will happen within 10,000 years. His view is based on several factors, including the empowerment of women and the advancement of contraceptive technology, the cost of having and raising children, the lack of genetic diversity in our species, climate change, our dominance of the world's natural economy, our dependence on a narrow range of plant-based food, and the evolutionary record, which shows that of the many varieties of hominins that came into being only one remains – us. Putting all of this together, Gee (2025) argues that our species reached its peak in the 1960s and that, like the Roman Empire, it is now on the path of decline and eventual fall.

Barring the implausible possibility that fertility rates in below replacement countries will go back above replacement level (Goldin, 2021; Hellstrand et al., 2021; Hwang, 2023; Wolf et al., 2011) and the rates in high fertility countries will not continue to decline (OECD, 2024), we agree with Gee that extinction will happen but we disagree on the timing: The work we present here suggests that humans will be extinct much faster. We disagree on the timing because in examining current world population estimates (2020 and 2025) by age (0-4, 5-9,…80-84, 85+) available from the U.S. Census Bureau (2025), we found that in trending out the (dramatic) five year rate of change in the Child-Adult Ratio (CAR, the number aged 0-4 divided by the number aged 15-44) over their respective five year periods, the human population would become extinct by 2394, only 369 years from now.

As an overview of the basis for these arguments, we first discuss current and future assumptions about the level of fertility. Second, we describe the data and method we use. Third, we describe an unrealistic but possible scenario in which fertility went to zero in 2024. We end the paper with a discussion that places our results in a deeper context.

**World Fertility Levels**

Estimating fertility (and mortality) for the world as a whole is not an easy task. Assembling the records from individual countries and attempting to harmonize them is a major task in itself. Adding to this task is the fact that some countries have "vital statistics" that are more or less complete, but many do not. Adair et al. (2023) report that of 194 countries, civil registration of births was only 77%, while completeness of vital statistics was only 63%. They found that the gap in completeness between civil registration and vital statistics for births is most pronounced in countries with lower civil registration completeness. Methodological tools (see, e.g., Popoff and Judson, 2003) and sample surveys are used to provide estimates where vital records are known or thought to be incomplete (see, e.g., Philippine Statistics Authority (PSA) and ICF, 2023). However, these are samples, subject not only to sampling error, but also to coverage, non-response, and measurement error (Swanson, 2013: 13-16).

Keeping the difficulty in mind, here are three current estimates of the world fertility in terms of the Total Fertility Rate (TFR): (1) 2.2 (Population Reference Bureau, 2025); (2) 2.25 (United Nations, 2024b); and (3) 2.29 (U.S. Census Bureau, 2025). Carney (2024) argues that the above replacement rates are too high and that it is likely that the world's fertility level is already below replacement.

The UN and Census Bureau fertility levels are present in world projections made by these organizations, as well as anyone else using these same data, respectively. In the case of the UN's projections, its medium variant has the 2100 population at 10,288,515 (UN, 2024a). Underlying this projection is a TFR that in 2024 is equal to 2.2464 (rounded above to 2.25) and by 2100 is 1.8390 (United Nations, 2024b). If either the TFR trend is extended to a longer-term projection or held constant at its 2100 level and extended to a longer-term projection, the medium variant of the UN's projection would ultimately yield a human population of zero. Similarly, underlying the U.S. Census Bureau's projections of the world population are TFRs of 2.31 in 2024 and 2.29 in 2024 and 2025, respectively (U.S. Census Bureau, 2025). These values are inherent in the Bureau's 2100 projection of the world population, 10.9 billion, which has an accompanying TFR of 1.81 (U.S. Census Bureau, 2025). Clearly, as is the case for the UN's projection, if this projection were extended further into the future, it also would yield a human population of zero.

**The Cohort Component Method of Population Projection**

Current and future world fertility, in the form of a TFR that overall is near replacement level, is inherently embedded in the fertility component of the approach employed in these projections - the Cohort Component method (CCM) (George et al., 2004; Smith, Tayman, and Swanson, 2013: 155-182; Yusuf, Martins and Swanson, 2014: 231-253). As its name suggests, the CCM requires the application of the components of population change – fertility, mortality, and migration to the age-gender structure at the projection's launch year. We begin with a discussion of the CCM, which also forms a point of departure for the forthcoming discussion of the Hamilton-Perry method, which we employ in this paper.

There are three *components of change* in a population: mortality, fertility, and migration. The overall growth or decline of a population is determined by the interplay among these three components. The exact nature of this interplay can be formalized in the *basic demographic equation*:

$$P_l - P_b = B - D + IM - OM, \qquad [1]$$

Where $P_l$ is the population at the end of the time period; $P_b$ is the population at the beginning of the time period; and B, D, IM, and OM are the number of births, deaths, in-migrants, and out-migrants during the time period, respectively. The difference between the number of births and the number of deaths is called *natural change* (B – D); it represents population growth coming from within the population itself. It may be either positive or negative, depending on whether births exceed deaths or deaths exceed births. The difference between the number of in-migrants and the number of out-migrants is called *net migration* (IM – OM); it represents population growth coming from the movement of people into and out of the area. It may be either positive or negative, depending on whether in-migrants exceed out-migrants or out-migrants exceed in-migrants.

The basic demographic equation can also be extended to apply to age groups, age-sex groups, and age-sex-race groups, as well as age-sex-ethnicity groups. This type of extension forms the logical basis of the and can be used to project a population into the future by age, age and sex, or by age, sex, and race. Once launched, these components (which are frequently modified as the projection moves into the future based on assumptions about their direction) are applied to the resulting age-gender structure at each cycle of the projection. At the world level, there is no migration, which eliminates the need for this component in world population projections.

**The Hamilton-Perry Method of Population Projection**

Instead of using the CCM approach, we employ its algebraic equivalent, the Hamilton-Perry (H-P) method (Baker et al., 2017: 251-252). Unlike the CCM approach, the H-P method does not apply the separate components of population change to the age-sex structure at the launch year. Instead, it computes cohort change ratios (CCRs) using two counts of the age-structure in question, typically five or ten years apart, which directly capture mortality and migration. The fertility component uses a "child-adult ratio" from the most recent age structure data or a "child-woman ratio" for a projection by gender. It is well-suited for generating a projection of the population of the world, per the framework found in Swanson et al. (2023):  (1) It corresponds to the dynamics by which a population moves forward in time; (2) there is information available relevant to these dynamics; (3) the time and resources needed to assemble relevant information and generate a projection is minimal; and (4) the information needed from the projection is generated by the H-P method.

The H-P method moves a population by age (and sex) from time t to time t+k using cohort-change ratios (CCRs) computed from data in the two most recent data points (e.g., censuses or estimates). It consists of two steps. The first uses existing data to develop CCRs, and the second applies the CCRs to the cohorts of the launch year population to move them into the future. The formula for the first step, the development of a CCR, is:

$$_nCCR_{x,i} = {_nP_{x,i,t}} / {_nP_{x-k,i,t-k}}, \quad [2]$$

where

$_nP_{x,i,t}$ is the population aged x to x+n in area i at the most recent census/estimate (t),

$_nP_{x-k,i,t-k}$ is the population aged x-k to x-k+n in area i at the 2$^{nd}$ most recent census/estimate (t-k),

k is the number of years between the most recent census/estimate at time t for area i and the census/estimate preceding it for area i at time t-k.

The basic formula for the second step, moving the cohorts of a population into the future, is:

$$_nP_{x+k,i,t+k} = (_nCCR_{x,i}) \times (_nP_{x,i,t}), \quad [3]$$

where

$_nP_{x+k,i,t+k}$ is the population aged x+k to x+k+n in area i at time t+k

Given the nature of the CCRs, they cannot be calculated for the youngest age group (e.g., ages 0-4 if it is a five-year projection cycle; 0-9 if it is a ten-year projection cycle), because this cohort came into existence after the census/estimate data collected at time t-k. To project the youngest age group, we use the "Child-Adult Ratio" (CAR), where the number in the youngest age group at time t is divided by the number of adults at time t who are of childbearing age (e.g., 15-44). It does not require any data beyond what is available in the census/estimate sets of successive data.

The CAR equation for projecting the population aged 0-4 is:

$$\text{Population 0-4: } {_5P_{0,t+k}} = ({_5P_{0,t}} / {_{30}P_{15,t}}) \times ({_{30}P_{15,t+k}}), \quad [4]$$

where

P is the population,

t is the year of the most recent census, and

t+k is the estimation year.

Projections of the oldest open-ended age group differ slightly from the H-P projections for the age groups beyond age 10 up to the oldest open-ended age group. If, for example, the final closed age group is 80-84, with 85+ as the terminal open-ended age group, then calculations for the $CCR_{i,x+}$ require the summation of the three oldest age groups to get the population age 75+ at time t-k:

$$_{\infty}CCR_{75,i,t} = {_{\infty}P_{85,i,t}} / {_{\infty}P_{75,i,t-k}} \qquad [5]$$

The formula for estimating the population of 85+ of area i for the year t+k is:

$$_{\infty}P_{85,i,t+k} = ({_{\infty}CCR_{75,i,t}}) \times ({_{\infty}P_{75,i,t}}). \qquad [6]$$

An issue that is found in the cohort change ratio for the terminal, open-ended age group (which in our case is 85 years and over) in a projection where migration is not a component of population change is that like the equivalent probability of survival in an abridged life table, deaths are not uniformly distributed within the interval (Chiang, 1984: Lahiri, 2018; Swanson, Bryan, and Chow, 2020). This issue tends to exaggerate the length of life for those aged 85 and over in an abridged life table and in an H-P projection. Because of it, we set the extinction of the human race 25 years after the point at which only those aged 85 and over are alive, which translates into the assumption that nobody lives beyond 110. While there are documented cases of people living beyond 110 years (Barbi et al., 2018), given the lack of social-economic and health support for them if there is nobody under the age of 110, we believe this is a reasonable approximation to the extinction of those aged 85 and over, as we discuss later

A disadvantage of the H-P method is that it can lead to unreasonably high estimates in rapidly growing places and unreasonably low projections in places experiencing population losses. Since the H-P and other extrapolation methods are based on population changes within a given area, it is essential to develop geographic boundaries that remain constant over time. Neither of these issues is a problem when the H-P method is applied to a country, much less the world as a whole.

Before turning to a discussion of the input data and the projection results, it is helpful to note that like the CCM, the H-P method is grounded in demographic theory. Barring unforeseeable catastrophes and other events that have very low probabilities of occurring (Taleb, 2010), the closer one comes to having accurate data embedded in a method that is grounded in demographic theory, the more accurate the projection method will likely be (Swanson et al., 2023).

**Input Data and Projection Results**

As input data, we employ data provided by the U.S Census Bureau via its International Data Base (2025) for 2019 and 2024, which are structured into five-year age groups (0-4, 5-9,…,75-79, 80-84) with 85+ being the terminal, open-ended age group. Table 1 shows the input data along with the 2019-2024 CCRs and the 2019 and 2024 CARs calculated from them. As expected, the CCRs are all less than 1.0, since they only reflect the mortality in moving from a younger to an older age group. They also show the typical mortality pattern by age, with mortality levels increasing with age, most dramatically after age 70. The drop in world fertility levels during the last five years is clearly evident, as the CAR declines by 7.5% between 2019 and 2024.

(TABLE 1 ABOUT HERE)

Table 2 shows the world population projection from 2024 to 2394, the year of extinction. It is based on two key assumptions: 1) the mortality rate based on the 2019-2024 CCRs is unchanged, and 2) the 7.5% drop in the fertility rates, as measured by the CAR, continues every five years into the future. Under this scenario, Homo sapiens will be completely extinct by 2394, only 369 years from now. The path to extinction is not linear due to the interaction between declining births and an increasingly older age structure. The population declines by -6.0% during the first 50 years of the projection horizon. That decline increases to -63.4% during the next 50 years. By 2124, the world's population is 2.8 billion, a drop of 5.3 billion people from 2024. The population drops by 86.9% during the next 50 years, reaching 0.36 billion in 2174.

(TABLE 2 ABOUT HERE)

Accompanying the changes in the total population are dramatic shifts in the age composition of the world population toward the elderly shown in Figure 1, which shows the share of the population in age groups under 20 years, 20-64 years, and 65 years and older. In 100 years (2124), the youngest age group is just 5.1% of the population, down from 32.4% in 2025. The share of the working ages (20-64) increases from 57.3% in 2024 to 62.5% in 2074, then starts a continuous decline. In 2124, that segment of the population (51.7%) is slightly below its share in 2024 (57.3%). Conversely, the share of the oldest age group (65+) rises dramatically from 10.3% in 2024 to 43.2% in 2124. By 2284, there are no more people aged under 20, 9.9% of the population is aged 20-64, and 90.1% are 65 years of age or older. The last of the population 20-64 is gone by 2359, 75 years after the end of the youngest age group. At that point, the remaining world population is all 65 years and older.

(FIGURE 1 ABOUT HERE)

This projection only considers current age-related survivorship probabilities, and the changes expected in age groups under decreasing fertility. Consider the absence of working-age people and the accompanying collapse of social and health services, as well as the economy as a whole. In this scenario, the survivorship rates (the CCRs) would most likely be lower than those used in this scenario In that case, the time to extinction would likely be quicker than reported here, particularly after 2359, when there would be nobody under the age of 65 to help support those aged 65 and over, and many of those alive would have difficulty supporting themselves and assisting with the support of others — a "Lord of the Flies" situation in reverse (Golding, 1954). So, considering the collapse of the family, social and health services, and the economy, it is plausible that the human population of the world could be gone within 334 years.

**Fertility Goes to Zero**

W also examined a scenario in which births worldwide ended in 2024. In this scenario, we used the same IBD world population data by age at two points in time, 2019 and 2024 (U. S. Census Bureau, 2025), we projected the world's population using the H-P method. Not surprisingly, the time to human extinction is much faster than the trended fertility scenario: 109 years compared to 369 years (see Table 3). The population decline is also not linear. It drops by 18.8% during the first 20 years of the projection horizon, reaching 6.5 billion. That decline increases to -30.9% during the next 20 years. By 2064, the world's population will be 4.5 billion, slightly more than half of the 2024 population of 8.1 billion. The population drops by 89.3% during the next 40 years, reaching 0.48 billion in 2104, 30 years before extinction.

(TABLE 3 ABOUT HERE)

Without births, the shifts in the age composition are very dramatic and show a more rapid increase toward the elderly (See Figure 2). The youngest age group, 32% of the population in 2024, becomes extinct by 2044. The share of the working ages (20-64) increases from 57.3% in 2024 to 80.6% in 2044, then starts a continuous decline. In 2069, that segment of the population (59.2%) is slightly above its share in 2024 (57.3%). The last of the population 20-64 is gone by 2089, 45 years after the end of the youngest age group. Conversely, the share of the oldest age group (65+) rises dramatically from 10.3% in 2024 to just under 50% by 2075. Starting in 2089, the remaining world population will be 65 years and older. Again, when there would be virtually nobody to help support those aged 65 and over, the human population of the world will likely be effectively gone by 2089, only 64 years from now.

(FIGURE 2 ABOUT HERE)

**Discussion**

Because the CAR is the key element in the projection, it is worthwhile to start the discussion with it, which is somewhat equivalent to the General Fertility Rate (GFR) but expanded to include males. For example, the population aged 0-4 at a point in time is primarily influenced by past births but it also is influenced by deaths and includes net migration. It is a ratio that measures the population per 1,000 women and men of childbearing age, which here we define as 15 to 44 years of age. It provides a good picture of current fertility or fertility within a given year and has the advantage of being easy to explain. Still, it is primarily driven by changes in the underlying age structure of the population, which means, for example, if people in their 40s have very few children and the population over 40 increases, the GFR will decrease if the underlying age-specific fertility rates (ASFRs) are not changing, specifically for those aged 40-44. This, however, is a fine point that does not substantively affect the projections. Similarly, if we changed the CAR so that its denominator ranged from age 10 to age 54 (as is the case with the projections done by the United Nations (2024a, 2024b), it would not substantively affect the H-P projections.

Speaking of the UN Projections, we used its 2020 and 2025 data (United Nations, 2025) to develop a trended 2020-2020 CAR in conjunction with a 2020-2025 H-P projection. We found results similar to the results we report based on the IDB data. Starting with a 2025 world population of approximately 8.232 billion, by 2285 it will be less than one million, at approximately 842,000; by 2325 it will be under 100,000 at approximately 70,000; by 2340 it will be less than 10,000 at approximately 7,100 and by 2360 less than 1,000 remain, all of which will be gone by 2430, only 29 years more than the extinction date (2399) found using the IDB data. In terms of an immediate zero births scenario, these same UN data show that by 2045 there would be nobody in the world under the age of 20 and its total population would be approximately 7.1 billion; by 2065, there would be nobody under the age of 40 and the population would be approximately 4.6 billion; by 2085, there would be nobody under the age of 60 and the total population would be approximately 2.4 billion; and by 2110, there would be nobody under the age of 85 and the total world population would consist of approximately 269 million people, all of whom would be 85 years and over. By 2135 all of them would be gone, only 110 years from now. Again, when there would be virtually nobody to help support those aged 65 and over, the human population of the world will likely be effectively gone by 2085, only 60 years from now.

Returning to the scenario in which births drop to zero in 2024, in the 2006 movie *Children of Men* (based on the 1992 book by P. D. James), male sperm counts have fallen so low that births worldwide have ended. This dystopian movie is set in England 20 years after the end of births, where a semblance of civil order remains. At the same time, other complex societies worldwide have already collapsed or are well into the process of collapse. While, as we noted earlier, it is unlikely that we will see an end to births in our lifetimes, it is not beyond the realm of possibility. Research indicates that male sperm counts are low worldwide, and the trend is further downward (Levine et al., 2023). Taken in conjunction with current low levels of birth rates due to the high cost of raising children and the cultural changes that empowered women to make their own reproductive decisions, fertility is already below the level needed to sustain populations in complex societies around the world.

All of the other events that could potentially be labeled as "extinction level" (Bostrom, 2022, 2009, 2013; Bostrom and Ćirković, 2011; Ćirković, Sandberg, and Bostrom, 2010; Posner, 2004), which include a giant asteroid strike, volcanic eruptions, plagues, wars involving nuclear, biological, or chemical weapons, rapid climate change, loss of food, are like the zero birth scenario: they get us to the same destination, small numbers of scattered survivors that will ultimately result in zero people – just sooner.

Barring the improbable possibility that fertility rates in below-replacement countries will go back above replacement level (Goldin, 2021; Hellstrand et al., 2021; Hwang, 2023; Wolf et al., 2011) and that in countries with rates at or above replacement levels will not continue to decline (OECD, 2024), we agree with Gee that extinction will happen, but disagree on the timing: The work we present here suggests that humans could be extinct in 370 years, well before 10,000 years have passed. We note that this is a deterministic projection, one that does not consider uncertainty (Taleb 2010), a factor only recently introduced to population projections in terms of the CCM and H-P methods (Alkema et al. 2015; Swanson and Tayman, forthcoming; Swanson, Tayman, and Cline, 2025; United Nations, 2024c).

Our scenario, which used the recent five-year trend in CARs to project future CARs, while not as extreme as the no birth scenario, implies that births will end 95 years before extinction in 2394. The CAR, which in 2024 is 0.183, falls below 0.050 by 2109, and is so low by 2299 that the population aged 0-5 is zero. Also, given the potential upheaval in social and economic conditions portended by lower fertility and population aging, it may be the case that our assumption of constant mortality rates by age (the CCRs found in Table 1) serves to underpredict future deaths, thus extending the extinction date. Additional study on the world population extinction question would benefit by evaluating alternative deterministic paths for the CCRs and CARs to reflect the impact of different trajectories for world fertility and mortality on the time to extinction.

We conclude with the fact that Gee's (2025) insight regarding the extinction of Homo sapiens was influenced by the work of Vollset, Goren, Yuan, et al. (2020) and Basten, Lutz, and Scherbov, (2013), scholars who recognized that world fertility rates are low and could go lower but nonetheless reported no results that ended in extinction.

Golding, W. (1954). Lord of the Flies. Faber and Faber. Demography (2021) 58(4):1373–1399 doi: 10.1215/00703370-9373618 © 202

Hellstrand, J., J. Nisén, V. Miranda, P. Fallesen, L. Dommermuth, and M. Myrskylä (2021). Not Just Later, but Fewer: Novel Trends in Cohort Fertility in the Nordic Countries. Demography 58(4):1373–1399. Doi:10.1215/00703370-9373618.

Hwang, J. (2023). Later, fewer, none? Recent trends in cohort fertility in South Korea. Demography 60(2): 563-582. Doi:10.1215/00703370-10585316.

James, P. D. (1993). The Children of Men. A.A. Knopf.

Jones, C., D. Zurell, and K. Wiesner (2023). Novel analytic methods for predicting extinctions in ecological networks. Ecological Monographs 94:e1601. doi:10.1002/ecm.1601

Lahiri, S. (2018). Survival Probabilities From 5-Year Cumulative Life Table Survival Ratios (Tx+5/Tx): Some Innovative Methodological Investigations. pp. 481-542 in A S. Rao and C. R. Rao (eds.) Handbook of Statistics, Volume 39. Elsevier.

Levine, H., N. Jørgensen, A. Martino-Andrade et al., (2023). Temporal trends in sperm count: A systematic review and meta-regression analysis of samples collected globally in the 20$^{th}$ and 21$^{st}$ centuries. Human Reproduction Update 29 (2): 157-176.

Liu, R., H. Ohashi; A. Hirata, L. Tang, T. Matsui, K. Terasaki, R. Furukawa; and N. Itsubo (2024). Predicting the global extinction risk for 6569 species by applying the life cycle impact assessment method to the impact of future land use changes. Sustainability 16, 5484. doi:10.3390/su61335484.

Lutz, W., and R. Qiang (2002). Determinants of human population growth. Philosophical Transactions of the Royal Society B: Biological Sciences 357 (1425): 1197.

Meadows, D. H., D. H. Meadows, J. Randers, and W. Behrens (1972). The limits to growth: A report to the club of Rome. Universe Press.
Swanson and Tayman 16

**Table 1. World population input data, CCRs and CARs**

| Age group | 2019 | 2024 | CCR |
|---|---:|---:|---:|
| 0 to 4 | 673,612,665 | 646,202,360 | |
| 5 to 9 | 667,710,912 | 667,775,896 | 0.99134 |
| 10 to 14 | 636,570,689 | 664,986,444 | 0.99592 |
| 15 to 19 | 601,991,281 | 633,797,040 | 0.99564 |
| 20 to 24 | 591,419,519 | 600,189,731 | 0.99701 |
| 25 to 29 | 602,625,686 | 589,354,041 | 0.99651 |
| 30 to 34 | 589,258,403 | 599,052,875 | 0.99407 |
| 35 to 39 | 537,712,957 | 584,415,902 | 0.99178 |
| 40 to 44 | 488,273,876 | 530,890,175 | 0.98731 |
| 45 to 49 | 475,747,986 | 480,137,829 | 0.98334 |
| 50 to 54 | 435,661,787 | 464,162,310 | 0.97565 |
| 55 to 59 | 369,556,392 | 420,078,544 | 0.96423 |
| 60 to 64 | 316,199,923 | 349,209,578 | 0.94494 |
| 65 to 69 | 257,401,498 | 289,929,534 | 0.91692 |
| 70 to 74 | 183,787,811 | 224,477,292 | 0.87209 |
| 75 to 79 | 128,656,826 | 147,938,707 | 0.80494 |
| 80 to 84 | 84,147,776 | 90,332,565 | 0.70212 |
| 85+ | 64,359,623 | 73,152,714 | 0.49259 |
| All Ages | 7,704,695,610 | 8,056,083,537 | |
| | | | Pct. chg. 2019-2024 |
| **CAR** | 0.19747 | 0.18266 | -7.5% |

Source: US Census Bureau (2025). Calculations by authors.

**Table 2. World population, 2024 - 2394**[a]

| Year | Population | Pct. chg. |
|---|---|---|
| 2024 | 8,056,083,537 | |
| 2074 | 7,569,619,136 | -6.0% |
| 2124 | 2,772,837,153 | -63.4% |
| 2174 | 362,750,313 | -86.9% |
| 2224 | 16,658,337 | -95.4% |
| 2274 | 273,161 | -98.4% |
| 2324 | 1,694 | -99.4% |
| 2374 | 4 | -99.7% |
| 2394 | 0 | -100.0% |

[a] Assumes 2019-2024 CCRs and percent decline in CARs.

**Table 3. World population, 2024 - 2134[a]**

| Year | Population |
|---|---|
| 2024 | 8,056,083,537 |
| 2044 | 6,537,988,033 |
| 2064 | 4,517,751,110 |
| 2084 | 2,290,687,469 |
| 2104 | 483,984,766 |
| 2124 | 28,494,504 |
| 2134 | 0 |

[a] Assumes 2019-2024 CCRs and no births after 2024.

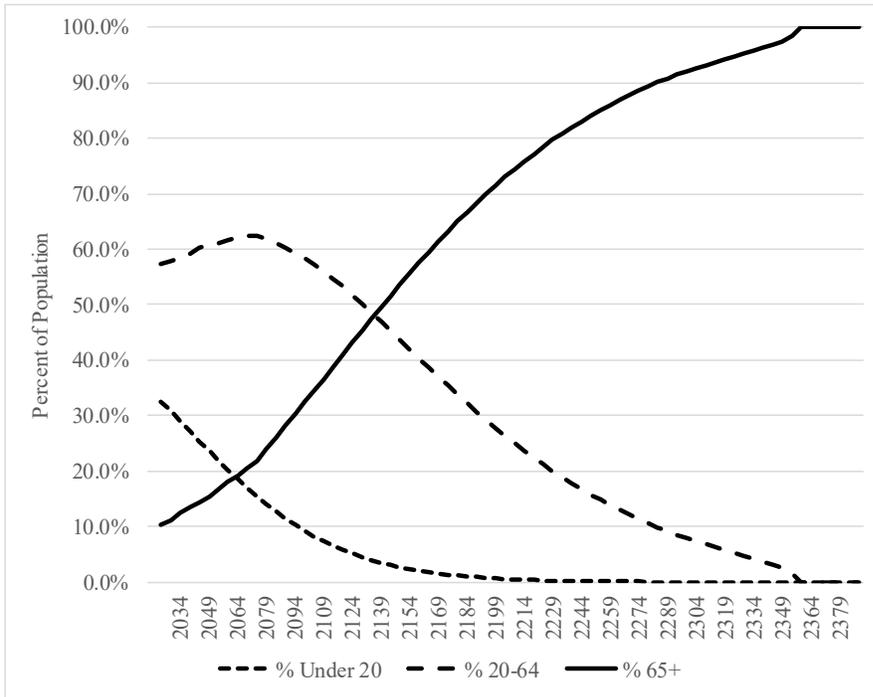

**Figure 1. World population age structure, 2024-2389, trended CAR scenario**

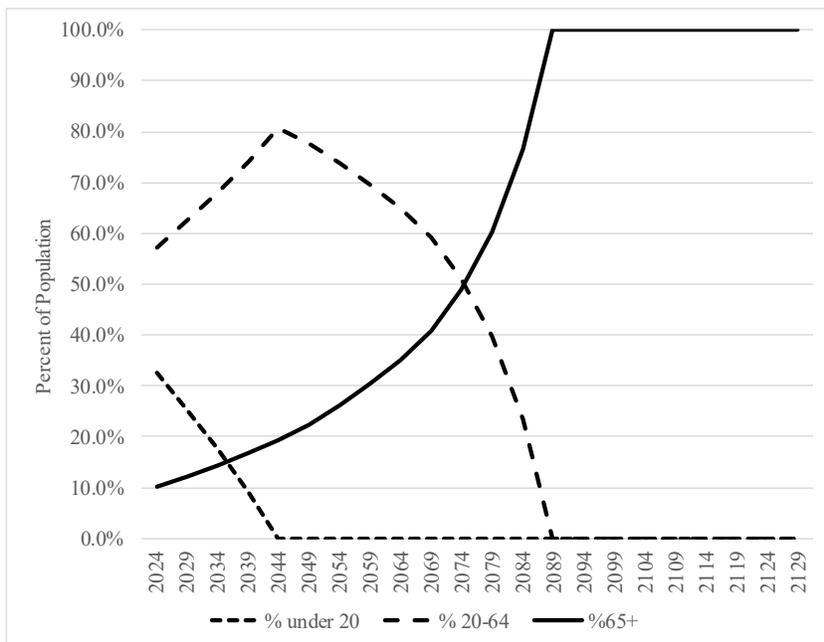

**Figure 2. World population age structure, 2024-2129, no births scenario**